\documentclass[11pt]{article}

\usepackage[T2A]{fontenc}
\usepackage[cp1251]{inputenc}
\usepackage[english,russian]{babel}
\usepackage{amsmath}
\usepackage{amsthm}
\usepackage{amssymb}
\usepackage{amscd}
\usepackage[dvips]{graphicx}
\usepackage{wrapfig}
\usepackage{titlesec}
\usepackage{color}
\usepackage{graphicx}
\usepackage{multicol}
\usepackage{floatflt}
\usepackage{indentfirst}

\theoremstyle{plain}
\newtheorem{Theorem}{Теорема}
\newtheorem{Lemma}{Лемма}
\newtheorem{Definition}{Определение}
\newtheorem{Statement}{Утверждение}

\theoremstyle{definition}
\newtheorem{Example}{Пример}
\newtheorem{Remark}{Замечание}

\newenvironment{Proof}{\par\noindent\textbf{Доказательство.}}{\hfill$\scriptstyle\blacksquare$\vspace{3mm}\par}

\renewcommand\Im{\mathop{\mathrm{Im}}}
\renewcommand\le{\leqslant}
\renewcommand\ge{\geqslant}
\renewcommand\phi{\varphi}

\newcommand\equivlogic{\Longleftrightarrow}
\newcommand\gterm[1]{\mathcal{#1}}
\newcommand\muld[1]{\mathbf{#1}}
\newcommand\abs[1]{\left|#1\right|}
\newcommand\res[1]{\mathop\mathrm{res}\limits_{#1}}

\newcommand\Laplace{\Delta}

\title{
    Достаточное условие несингулярности дискретного конечнозонного при одной энергии двумерного оператора Шредингера на квад-графе.
    \thanks{Работа выполнена при поддержке гранта Правительства Российской Федерации 2010-220-01-077}
}
\author{
    Б.\,О.\,Василевский \thanks{МГУ им. М.\,В.~Ломоносова, email: vasilevskiy.boris@gmail.com}
}
\date{8 июля 2013}

\begin{document}

\maketitle
\begin{abstract}
    Используя конечнозонный подход, строится двумерный дискретный оператор Шредингера на квад-графе, планарном графе, у которого каждая грань~--- четырехугольник. После этого предлагается определение несингулярности этого оператора как положительности всех его коэффициентов и к этому приводятся достаточные условия на спектральную кривую и сам граф.
\end{abstract}


\subsection{Введение}
    В теории интегрируемых систем математической физики важную роль играет конечнозонный подход. В непрерывном случае хотелось бы упомянуть работу Дубровина, Кричевера и Новикова~\cite{periodicShredinger}, в которой они показали интегрируемость двумерного стационарного конечнозонного оператора Шрёдингера при фиксированной энергии. Не только чисто теоретический интерес вызывает задача построения интегрируемых дискретизаций этого оператора. Интегрируемая гиперболическая дискретизация на квадратной решетке (построено обратное спектральное преобразование в периодическом случае) была найдена И.\,М.\,Кричевером~\cite{giperbolDiscr}. Далее, в статье А.\,Доливы, П.\,Гриневича, М.\,Нишпровски и П.\,Сантини~\cite{4authors} была получена эллиптическая дискретизация из специальной редукции гиперболической дискретизации. Эта редукция в терминах спектральных данных оказалась очень похожа на редукцию в работах Веселова и Новикова~\cite{finitShredinger} для непрерывного случая. В частности, на спектральной кривой требуется наличие голоморфной инволюции с двумя неподвижными точками.

    Случаи двух и нуля неподвижных точек у голоморфной инволюции на римановой поверхности являются наиболее интересными, согласно Д.\,Фэю~\cite{Fay}. Как показали Кричевер и Грушевский~\cite{KG}, конечнозонные решения, построенные в~\cite{4authors}, являются решениями специального вида. Решения общего положения отвечают спектральным кривым, у которых инволюция не имеет неподвижных точек. Но вслед за~\cite{4authors} мы будем рассматривать инволюцию именно c двумя неподвижными точками.

    Хорошим обобщением квадратной решетки является квад-граф, у которого каждая грань по построению является четырехугольником. Здесь хотелось бы отметить работу А.\,Бобенко, К.\,Меркат, Ю\,Сурис~\cite{discran}. В одной из ее глав обсуждается вложение квад-графа в комплексную плоскость, при котором каждая грань переходит в параллелограмм, и интегрируемость возникающего при таком вложении оператора Коши-Римана в смысле <<3D-совместности>>. При этом случай положительных весов в точности соответствует уже ромбовидному вложению и является самым интересным.

    Здесь мы применяем конечнозонный подход для построения дискретного оператора Шредингера на квад-графе. Обсуждается идея о том, что правильным определением несингулярности дискретного оператора является как раз его положительность. Кроме того, приводятся условия на обобщенные спектральные данные и квад-граф, достаточные для положительности весов построенных операторов. Примечательно, что для спектральных данных работают условия, абсолютно аналогичные условиям из~\cite{finitShredinger} для выделения несингулярных и строго положительных потенциалов двумерного оператора Шредингера в непрерывном случае.

\subsection{Дискретные комплексы, квад-графы и комплексный анализ}
    Рассмотрим двумерный дискретный подкомплекс $\Omega_\gterm{D}$ $d$-мерной квадратной решетки $\mathbb{Z}^d$ для произвольного $d \ge 2$. Каждая грань $\Omega_\gterm{D}$ является двумерным единичным квадратом. Потребуем, чтобы подкомплекс укладывался $\mathbb{C}$ без самопересечений, то есть представлялся в виде планарного графа $\gterm{D}$. Обратное отображение отправляет вершины графа в узлы данной решетки $\muld{n}: V(\gterm{D}) \to \mathbb{Z}^d$. Каждая грань $\gterm{D}$ является четырехугольником, сам граф является двудольным.

    Возьмем вершины одной доли и соединим ребрами те из них, которые лежат в одной грани. Полученный граф обозначим через $\gterm{G}$, а построенный аналогично по другой доле --- $\gterm{G}^*$. Несложно видеть, что $V(\gterm{D}) = V(\gterm{G}) \sqcup V(\gterm{G}^*)$. Кроме того, для любой грани $\gterm{D}$ одна из ее диагоналей $e$ является ребром в $\gterm{D}$, а другая $e^*$ --- ребром в $\gterm{D}^*$. Наконец, $\gterm{G}$ и $\gterm{G}^*$ являются двойственными. Правильно выбрав поверхность $\Omega_\gterm{D}$, можно получить произвольный (планарный) граф $\gterm{G}$.

    Напомним некоторые определения из линейной теории дискретного комплексного анализа. Более подробное изложение можно найти например в~\cite{discran}. Пусть на ребрах графа $\gterm{G}$ определена комплекснозначная функция $\nu: E(\gterm{G}) \to \mathbb{C}$. Рассмотрим оператор $\Delta$, действующий на функциях $f: V(\gterm{G}) \to \mathbb{C}$ по формуле
    \begin{gather}\label{eqGenericLaplaceDef}
        (\Delta f)(x_0) = \sum\limits_{x \sim x_0} \nu(x_0, x)(f(x) - f(x_0)),
    \end{gather}
    где суммирование проходит по всем соседним с $x_0$ вершинам в графе $\gterm{G}$. Назовем этот оператор Лапласианом, соответствующим весовой функции $\nu$. Дискретной гармонической (относительно весовой функции $\nu$) назовем функцию функцию $f: V(\gterm{G}) \to \mathbb{C}$, для которой выполняется $\Delta f = 0$.

    Продолжим весовую функцию на ребра $e^* \in E(\gterm{G}^*)$ по формуле $\nu(e^*) = 1 / \nu(e)$, где $e \in E(\gterm{G})$ --- вторая диагональ соответствующего четырехугольника.
    \begin{figure}[ht]
        \centering
        \includegraphics{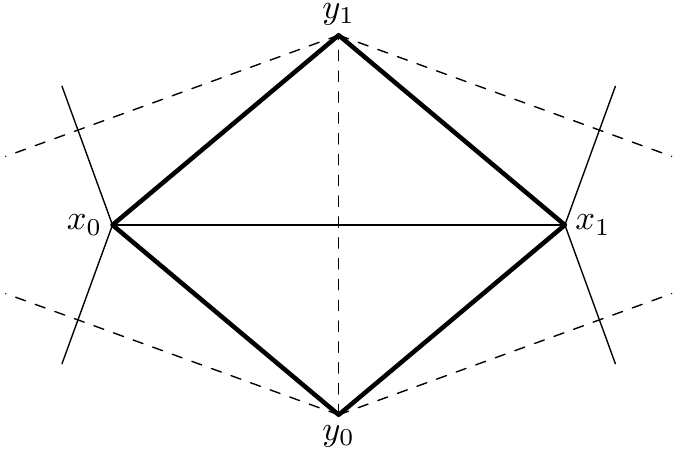}\\
        \caption{Грань $(x_0, y_0, x_1, y_1) \in F(\gterm{D})$. Тонкими линиями нарисованы ребра $\gterm{G}$, пунктиром~--- ребра $\gterm{G}^*$.}
        \label{picDoubleGraphFace}
    \end{figure}
    Голоморфные функции живут уже на вершинах $\gterm{D}$. Функция $f: V(\gterm{D}) \to \mathbb{C}$ называется дискретной голоморфной относительно весовой функции $\nu$, если для любой положительно ориентированной грани $(x_0, y_0, x_1, y_1) \in F(\gterm{D})$ (см. рисунок~\ref{picDoubleGraphFace}) выполняются дискретные уравнения Коши-Римана
    $$
        \frac{f(y_1) - f(y_0)}{f(x_1) - f(x_0)} = i\nu(x_0, x_1) = -\frac{1}{i\nu(y_0, y_1)}.
    $$

    Несложным вычислением проверяется, что ограничение дискретной голоморфной функции на любую из долей $V(\gterm{G})$, $V(\gterm{G}^*)$ является гармонической. Обратно, по любой гармонической функции на $V(\gterm{G})$ строится дискретная голоморфная на $V(\gterm{D})$, однозначная с точностью до прибавления константы на $V(\gterm{G}^*)$.

    Ребру из $\gterm{D}$ удобно присвоить в качестве метки тот координатный вектор $\mathbb{Z}^d$, в который переходит это ребро, а также ориентировать в сторону увеличения координаты.

    Отметим, что отображение $P$ любой квазикристаллической решетки в $\mathbb{Z}^d$, описанное в п. 3~\cite{discran}, имеет ровно такой же смысл, что и $\muld{n}$.

\subsection{Многоточечная волновая функция и дискретные уравнения Коши-Римана}
    Пусть дан планарный граф $\gterm{G}$. Построим по нему квад-граф $\gterm{D}$. Пусть существует вложение $\gterm{D}$ в целочисленную решетку $n: V(\gterm{D}) \to \mathbb{Z}^d$ для некоторого произвольного $d \ge 2$. Используя конечнозонный подход, мы построим по этому вложению весовую функцию $\nu$, через которую выписываются уравнения Коши-Римана на квад-графе $\gterm{D}$. Весовая функция, в свою очередь, определяется через обобщенные спектральные данные, о которых сейчас пойдет речь.

    Описанная ниже конструкция спектральных данных обобщает построения, сделанные в~\cite{giperbolDiscr} и в~\cite{4authors}. А именно, в указанных работах рассматривается случай квадратной решетки на плоскости и $d = 2$.

    Рассмотрим компактную, регулярную риманову поверхность $\Gamma$ рода $g$. Пусть на ней имеются следующие точки.
    \begin{itemize}
        \item Фиксированная точка $R_1$ на $\Gamma$ для нормировки волновой функции.
        \item Дивизор общего положения $\gamma_1, \dots, \gamma_g$.
        \item Коллекция из $d$ пар выделенных точек $A^+_1, A^-_1, A^+_2, A^-_2, \dots, A^+_d, A^-_d$. Все точки попарно различны.
    \end{itemize}
    По теореме Римана-Роха, для любого целочисленного вектора
    $$
        \muld{n} = (n_1, n_2, \dots, n_d) \in \mathbb{Z}^d
    $$
    существует единственная функция $\Psi(\muld{n}; \gamma)$, $\gamma \in \Gamma$, со следующими свойствами.
    \begin{enumerate}
        \item При каждом $\muld{n}$ функция $\Psi$ является мероморфной от $\gamma$.
        \item $\Psi$ имеет полюса не более первого порядка в точках $\gamma_1, \gamma_2, \dots, \gamma_g$.
        \item Для каждого $j = 1, 2, \dots, d$, $\Psi$ имеет полюс не более чем $n_j$ порядка в точке $A^+_j$ и нуль по крайней мере $n_j$ порядка в точке $A^-_j$.
        \item Выполняется условие нормировки $\Psi(\muld{n}; R_1) \equiv 1$.
    \end{enumerate}
    Функцию $\Psi$ называют волновой. Она естественно переводится на граф $\gterm{D}$:
    $$
        \Psi(p, \gamma) = \Psi(\muld{n}(p); \gamma), \quad p \in V(\gterm{D}).
    $$

\begin{figure}[h]
        \centering
        \includegraphics{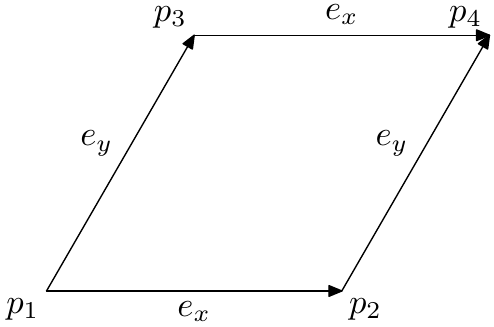}\\
        \caption{Грань $(p_1, p_2, p_4, p_3) \in F(\gterm{D})$.}
        \label{picGenericFace1}
    \end{figure}

    Рассмотрим произвольную (положительно ориентированную) грань $(p_1, p_2, p_4, p_3) \in F(\gterm{D})$. Пусть ребра этой грани были ориентированы в сторону вершин с большими номерами (см. рисунок~\ref{picGenericFace1}). Пусть ребро $(p_1, p_2)$ имеет метку $e_x$, а ребро $(p_1, p_3)$~--- метку $e_y$, $1 \le x, y \le d$, $x \ne y$.

    Заметим, что при данных общего положения у $\Psi(p_2, \gamma)$ полюс в $A^+_x$ на единицу большего порядка по сравнению с $\Psi(p_1, \gamma)$, а порядки остальных полюсов у них совпадают. Аналогично отличие видно между $\Psi(p_3, \gamma)$ и $\Psi(p_1, \gamma)$ по отношению к точке $A^\pm_y$. При переходе от $p_2$ или $p_3$ к $p_4$ наблюдается сходная картина. Сделанные замечания приводят нас к равенству
    \begin{gather}\label{eq4pointGeneric}
        \Psi(p_4, \gamma) + \alpha_1(p_1, p_2)\Psi(p_2, \gamma) + \alpha_2(p_1, p_3)\Psi(p_3, \gamma) + \alpha_3(p_1, p_4)\Psi(p_1, \gamma) = 0,
    \end{gather}
    где коэффициенты $\alpha_j$ не зависят от $\gamma$ и определяются по следующим формулам:
    \begin{gather}
        \alpha_1(p_1, p_2) = -\lim\limits_{\gamma \to A^+_x} \frac{\Psi(p_4, \gamma)}{\Psi(p_2, \gamma)}, \label{eqAlpha1} \\
        \alpha_2(p_1, p_3) = -\lim\limits_{\gamma \to A^+_y} \frac{\Psi(p_4, \gamma)}{\Psi(p_3, \gamma)}, \label{eqAlpha2} \\
        \alpha_3(p_1, p_4) = -1 - \alpha_1(p_1, p_2) - \alpha_2(p_1, p_3).
    \end{gather}
    Действительно, сумма \eqref{eq4pointGeneric} удовлетворяет всем условиям для $\Psi(p, \gamma)$ за исключением того, что в $R_1$ она обращается в нуль. Тогда по теореме Римана-Роха сумма равна нулю при любом $\gamma \in \Gamma$.

    Проделанные к этому моменту алгебро-геометрические построения аналогичны~\cite{giperbolDiscr}.

    Для того, чтобы уравнение~\eqref{eq4pointGeneric} привести к виду дискретного Коши-Римана, нам потребуются дополнительные условия на обобщенные данные рассеяния.
    \begin{Lemma}\label{lemmaSigmaOmega}
        Пусть на $\Gamma$ существует голоморфная инволюция $\sigma$ c двумя неподвижными точками $R_+ = R_1$ и $R_-$. Пусть спектральные данные обладают следующей симметрией.
        \begin{enumerate}
            \item Для $j = 1, \dots, d$ выполняется $\sigma A^+_j = A^-_j$.
            \item Существует мероморфный дифференциал $\Omega$ с двумя полюсами первого порядка в неподвижных точках $R_+$, $R_-$ и $2g$ нулями в $\gamma_1, \dots, \gamma_g$, $\sigma\gamma_1, \dots, \sigma\gamma_g$.
        \end{enumerate}
        Тогда
        \begin{gather}
            \Psi(p, R_-) = (-1)^{\abs{\muld{n}(p)}}, \label{eqRminusValue} \\
            \alpha_1(x_0, y_0) = -\alpha_2(x_0, y_1), \quad \alpha_3(x_0, x_1) = -1,
        \end{gather}
        где $\abs{\muld{n}} = n_1 + n_2 + \dots + n_d$.
    \end{Lemma}
    \begin{Proof}
        Ход доказательства сходен с~\cite{4authors}. По аналогии с~\cite{finitShredinger}, рассмотрим мероморфный дифференциал
        $$
            \tilde \Omega(p, \gamma) = \Psi(p, \gamma)\Psi(p, \sigma\gamma)\Omega(\gamma).
        $$
        У него два простых полюса в точках $R_+$, $R_-$ и никаких других особенностей. Следовательно, вычеты в них противоположны, как и у $\Omega$. Поскольку $\Psi(p, R_+) = 1$, то
        \begin{gather}\label{eqTempDiscreteRminus}
            \Psi^2(p, R_-) \equiv 1.
        \end{gather}
        Докажем, что для любого $\muld{n} \in \mathbb{Z}^d$ выполняется $\Psi(\muld{n}; R_-) = (-1)^{\abs{\muld{n}}}$. По единственности и выполнению всех свойств волновой функции, $\Psi(\muld{0}, \gamma) \equiv 1$, поэтому при $\muld{n} = \muld{0}$ равенство верно. Зададимся произвольным $1 \le j \le d$ и рассмотрим $\Psi(\muld{n} + \muld{e}_j; \gamma)$ как функцию от $A^+_j$ (вспомним про $A^-_j = \sigma A^+_j$). Из явной формулы для $\Psi$ вытекает, что $\Psi(\muld{n}, \gamma)$ является мероморфной от $A^+_j$. Пусть $A^+_j$ расположена в окрестности $R_-$. Выберем в этой окрестности такой локальный параметр $z$, чтобы $z(R_-) = 0$ и $\sigma z = -z$. Тогда
        $$
            \Psi(\muld{n} + \muld{e}_j; \gamma) = \frac{z + A^+_j}{z - A^+_j} \Psi(\muld{n}; \gamma) + o(A^+_j),
        $$
        откуда по~\eqref{eqTempDiscreteRminus} и непрерывности $\Psi(\muld{n} + \muld{e}_j; \gamma)$ получаем $\Psi(\muld{n} + \muld{e}_j; R_-) = -\Psi(\muld{n}; R_-)$ при любом расположении $A^+_j$. Из проведенных рассуждений легко следует искомое равенство.

        Подставляя~\eqref{eqRminusValue} в~\eqref{eq4pointGeneric}, получаем второе утверждение леммы.
    \end{Proof}
    \begin{Remark}
        Будем считать нормировку дифференциала $\Omega$ такой, что его вычеты в точках $R_+$, $R_-$ равны соответственно $\tfrac12$, $-\tfrac12$.
    \end{Remark}
    \begin{Remark}
        Условия леммы накладывают довольно сильные ограничения на спектральные данные.
    \end{Remark}

    Хочется отметить, что многоточечная волновая функция вместе с уравнением~\eqref{eq4pointGeneric} были построены еще в~\cite{discrPsiFunc} при более общих предположениях. В работе они используются для построения дискретного аналога решетки Дарбу--Егорова, размерность которой равна половине от количества неподвижных точек инволюции $\sigma$.

    Мы готовы определить весовую функцию $\nu: E(\gterm{G}) \sqcup E(\gterm{G}^*) \to \mathbb{C}$. Пусть условия леммы~\ref{lemmaSigmaOmega} выполнены. Вернемся к рассмотрению произвольной грани с описанной нумерацией вершин $(p_1, p_2, p_4, p_3) \in F(\gterm{D})$. Положим по определению
    \begin{gather}
        \nu(p_1, p_4) = \frac{1}{i \alpha_1(p_1, p_2)} = \frac{i}{\alpha_2(p_1, p_3)} \label{eqNu}, \\
        \nu(p_2, p_3) = \frac{1}{\nu(p_1, p_4)}.
    \end{gather}
    Перепишем равенство~\eqref{eq4pointGeneric} в терминах весовой функции
    $$
        \frac{\Psi(p_3, \gamma) - \Psi(p_2, \gamma)}{\Psi(p_4, \gamma) - \Psi(p_1, \gamma)} = i\nu(p_1, p_4) = -\frac{1}{i\nu(p_2, p_3)}.
    $$
    Рассмотрением всех 4 случаев взаимной ориентации ребер несложно показать, что данное равенство верно и без условия на ориентацию ребер. Можно сказать, что выполнением именно этого свойства продиктовано правило продолжения весовой функции с ребер $\gterm{G}$ на ребра двойственного графа $\nu(e^*) = 1/\nu(e)$.

    Таким образом, для произвольной грани ${(x_0, y_0, x_1, y_1) \in F(\gterm{D})}$ (рисунок~\ref{picDoubleGraphFace}) выполняется
    \begin{gather}\label{eqCauchyRiemannPsi}
        \frac{\Psi(y_1, \gamma) - \Psi(y_0, \gamma)}{\Psi(x_1, \gamma) - \Psi(x_0, \gamma)} = i\nu(x_0, x_1) = -\frac{1}{i\nu(y_0, y_1)}.
    \end{gather}
    По определению, волновая функция $\Psi(p, \gamma)$ является дискретной голоморфной на графе $\gterm{D}$ с весовой функцией $\nu$ при любом фиксированном $\gamma \in \Gamma$. Ограничение $\Psi(p, \gamma)$ на каждый из $\gterm{G}$, $\gterm{G}^*$ дает гармоническую функцию
    \begin{gather}\label{eqLaplacePsi}
        (L_x \Psi)(x_0, \gamma) = \sum\limits_{x \sim x_0} \nu(x_0, x)(\Psi(x, \gamma) - \Psi(x_0, \gamma)) = 0, \\
        (L_y \Psi)(y_0, \gamma) = \sum\limits_{y \sim y_0} \nu(y_0, y)(\Psi(y, \gamma) - \Psi(y_0, \gamma)) = 0.
    \end{gather}
    По умолчанию под $L$ мы будем подразумевать оператор $L_x$.

    В работе~\cite{discran} неоднократно отмечается особый интерес случая положительной весовой функции. Следующая лемма формулирует условия на обобщенные спектральные данные, достаточные для вещественности получаемой $\nu$. Случай положительных весов будет обсуждаться в разделе~\ref{sectionPositiveWeightSufficientCondition}.
    \begin{Lemma}\label{lemmaTau}
        Пусть выполнены условия леммы~\ref{lemmaSigmaOmega}. Пусть на $\Gamma$ существует антиголоморфная инволюция $\tau$ со следующими свойствами.
        \begin{enumerate}
            \item $\tau$ коммутирует с $\sigma$.
            \item $\tau R_+ = R_-$.
            \item Точки $A^+_1, A^-_1, \dots, A^+_d, A^-_d$ являются неподвижными для $\tau$.
            \item Дивизор $\gamma_1, \dots, \gamma_g$ переходит в себя под действием $\tau$ (но любая точка дивизора может переходить в отличную от себя).
        \end{enumerate}
        Тогда весовая функция $\nu$ принимает вещественные значения и выполняется
        \begin{gather}\label{eqRealityCondition}
            \Psi(p, \tau\gamma) = (-1)^{\abs{\muld{n}(p)}}\overline{\Psi(p, \gamma)}.
        \end{gather}
    \end{Lemma}
    \begin{Proof}
        Рассмотрим функцию $(-1)^{\abs{\muld{n}(p)}}\overline{\Psi(p, \tau\gamma)}$. Она является мероморфной и удовлетворяет всем условиям волновой функции, включая условие нормировки. По единственности она равна $\Psi(p, \gamma)$, откуда получается искомое равенство~\eqref{eqRealityCondition}. Таким образом, в неподвижных точках инволюции $\tau$ волновая функция принимает вещественные значения на одной из долей $\gterm{D}$ и чисто мнимые~--- на другой. Для доказательства вещественности $\nu = i/\alpha_2$ достаточно заметить, что в определении $\alpha_2$~\eqref{eqAlpha2} вершины в числителе и знаменателе относятся к разным долям.
    \end{Proof}

    Проделанные к текущему моменту алгебро-геометрические построения аналогичны~\cite{4authors}. Инволюция $\tau$ также используется и в~\cite{discrPsiFunc} как условие вещественности решетки.

    \begin{Example}
        Покажем, что квазикристаллический случай полностью описывается спектральными данными, в которых $\Gamma$ является сферой Римана. Будем считать, что кроме вложения квад-графа $\muld{n}$ в $d$-мерную решетку задан набор попарно линейно независимых комплексных чисел $\{ \alpha_1, \dots, \alpha_d \}$. Каждому координатному вектору $\muld{e}_j \in \mathbb{Z}^d$ поставим в соответствие число $\alpha_j \in \mathbb{C}$. Этим действием мы задаем квазикристаллическое вложение квад-графа в комплексную плоскость, описанное в примере п.~3~\cite{discran}.

        Пусть $\Gamma = \overline{\mathbb{C}}$, инволюция $\sigma$ задается центральной симметрией $\sigma z = -z$ с неподвижными точками $R_- = 0$, $R_+ = \infty$, дифференциал $\Omega = -dz/2z$, гамма-дивизор пустой. В качестве выделенных точек $A^\pm_j$ на $\Gamma$ возьмем $\pm\alpha_j$. Волновая функция на сфере --- это дискретная экспонента
        $$
            \Psi(\muld{n}; z) = \prod\limits_{j = 1}^{d} \left(\frac{z + \alpha_j}{z - \alpha_j}\right)^{n_j}.
        $$
        Несложно проверить, что построенная по $\Psi$ весовая функция равна
        $$
            \nu(p_2, p_3) = \frac{1}{\nu(p_1, p_4)} = i\frac{\alpha_y - \alpha_x}{\alpha_y + \alpha_x}.
        $$
        Она совпадает с построенной по квазикристаллическому вложению: значение $i\nu$ на любом параллелограмме равно отношению его диагоналей. В~\cite{discran} также доказано, что это равносильно интегрируемости уравнения Коши-Римана в смысле <<3D-совместности>>.

        В случае вещественных весов из последней формулы следует, что все $\alpha_j$ равны между собой по абсолютной величине: $\abs{\alpha_j}^2 = C^2$. Определим антиголоморфную инволюцию $\tau z = C / \overline{z}$, тогда все $\alpha_j$ будут для нее неподвижными точками и условия леммы~\ref{lemmaTau} будут выполнены. Таким образом, в квазикристаллическом случае достаточное условие вещественности является еще и необходимым.
    \end{Example}

\subsection{Достаточное условие несингулярности оператора Лапласа}
\label{sectionPositiveWeightSufficientCondition}
    Рассмотрим связный планарный граф $\gterm{G}$, вложенный в $\mathbb{C}$. Пусть на его ребрах определена весовая функция $\nu: E(\gterm{G}) \to \mathbb{C}$. Выберем в $\gterm{G}$ простой цикл $p = (x_1, x_2, \dots, x_n, x_1)$, он делит $\mathbb{C}$ на две части. Выберем одну из частей и назовем все вершины $x \notin p$, лежащие в этой части, внутренними по отношению к $p$. Граничными будем называть вершины цикла $p$. По построению, соседями внутренних вершин могут быть только внутренние и граничные.

    Рассмотрим класс $K$ комплекснозначных функций, определенных на всех внутренних и граничных вершинах. Введем оператор Лапласа $\Laplace$ на таких функциях: результат на любой внутренней вершине определяется по стандартной формуле~\eqref{eqGenericLaplaceDef}, а значение результата на границе равно 0.

    Теперь рассмотрим следующую краевую задачу. Пусть в вершинах пути $p$ заданы некоторые граничные условия. Необходимо найти все $f \in K$, удовлетворяющие граничным условиям и обнуляющие оператор Лапласа: $\Laplace f = 0$.
    \begin{Statement}
        Пусть весовая функция $\nu$ принимает только положительные значения. Тогда для любого простого пути $p$ и для любых граничных условий существует не более одного решения краевой задачи $f \in K$.
    \end{Statement}
    \begin{Proof}
        Перепишем условие гармоничности функции во внутренних вершинах $x_0$
        $$
            \sum\limits_{x \sim x_0} \nu(x_0, x) f(x_0) = \sum\limits_{x \sim x_0} \nu(x_0, x) f(x).
        $$
        Отсюда следует дискретный аналог принципа максимума
        $$
            \abs{f(x_0)} \le \max\limits_{x \sim x_0} \abs{f(x)}.
        $$
        Действительно, в противном случае левая часть равенства будет больше правой по абсолютной величине.

        Если в $K$ найдется два различных решения данной краевой задачи $f_1$, $f_2$, то $(f_1 - f_2)$ является ненулевым решением $\Laplace f = 0$ при нулевых граничных условиях, что противоречит принципу максимума.
    \end{Proof}
    В связи с этим логично ввести следующее определение несингулярности дискретного оператора. Вместо положительности удобнее говорить о значениях одного знака.
    \begin{Definition}
        Оператор $\Laplace$ называется несингулярным, если соответствующая ему весовая функция $\nu : E(\gterm{G}) \to \mathbb{R}$ принимает только значения одного знака.
    \end{Definition}
    Далее в этом разделе мы будем искать достаточные условия положительности либо отрицательности весовой функции, построенной по обобщенным спектральным данным.

    Пусть выполнены условия лемм~\ref{lemmaSigmaOmega} и~\ref{lemmaTau}. Пусть также $\Gamma$ является M-кривой. Это означает, что антиголоморфная инволюция $\tau$ имеет $g + 1$ неподвижный овал $a_1, a_2, \dots, a_g, c$. Заметим, что инволюция $\sigma$ переводит неподвижные точки $\tau$ в неподвижные благодаря условию коммутации: $\gamma = \tau\gamma \Rightarrow \sigma\gamma = \sigma\tau\gamma = \tau\sigma\gamma$. Из непрерывности следует, что $\sigma$ овалы переводит целиком в овалы.

\begin{figure}[h]
        \centering
        \includegraphics{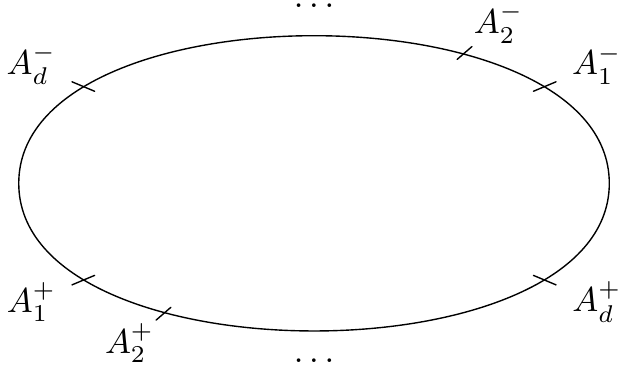}\\
        \caption{Расположение выделенных точек на овале $c$}
        \label{picPointAonOval}
    \end{figure}

    По определению $\tau$, выделенные точки $A^\pm_1, A^\pm_2, \dots, A^\pm_d$ являются неподвижными. Потребуем, чтобы все они попадали на один овал: $A^\pm_j \in c$, $j = 1, 2, \dots, d$. Отсюда $\sigma c = c$. Из сохранения ориентации инволюцией $\sigma$ вытекает справедливость следующего утверждения.
    \begin{Statement}
        При подходящей нумерации точек $A^\pm_j$, $j = 1, 2, \dots, d$, они располагаются на овале $c$ в порядке (см. рисунок~\ref{picPointAonOval})
        $$
            A^+_1, A^+_2, \dots, A^+_d, A^-_1, A^-_2, \dots, A^-_d.
        $$
    \end{Statement}
    \begin{Remark}\label{remarkProperPointsEnumeration}
        Будем считать, что именно такая нумерация была выбрана изначально и паре $A^\pm_j$ соответствует $j$-й координатный вектор $\muld{e}_j$.
    \end{Remark}

    Для каждой точки $A^+_j$, $j = 1, 2, \dots, d$, выберем такой локальный параметр $z_j$, чтобы в окрестности $A^+_j$ все точки овала $c$ были вещественными: $\Im z_j = 0$. Это в точности такие параметры, в которых $\tau z = \bar z$. Для любой вершины $p$ квад-графа рассмотрим разложение волновой функции $\Psi(p, \gamma)$ и двойственной ей $\Psi^+(p, \gamma) = \Psi(p, \sigma\gamma)$ в выбранных окрестностях:
    \begin{gather}
        \Psi(p, z_j) = r_j(p) z_j^{-\muld{n}_j(p)} + O\left( z_j^{-\muld{n}_j(p) + 1} \right), \label{eqMainKoef} \\
        \Psi^+(p, z_j) = r^+_j(p) z_j^{\muld{n}_j(p)} + O\left(z_j^{\muld{n}_j(p) + 1}\right), \label{eqMainKoefPlus}
    \end{gather}
    где $r_j(p)$, $r^+_j(p)$~--- ненулевые комплексные числа для всех $j = 1, 2, \dots, d$ и $p \in V(\gterm{D})$.
    \begin{Remark}\label{remarkLocParamForOtherPoints}
        В окрестности каждой из $A^-_j$, $j = 1, 2, \dots, d$, в качестве локального параметра можно взять $\sigma z_j$, $\tau\sigma z_j = \overline{\sigma z_j}$. Поскольку $\Psi(p, \sigma\gamma) = \Psi^+(p, \gamma)$, то~\eqref{eqMainKoefPlus} эквивалентно разложению $\Psi(p, \gamma)$ в окрестности $A^-_j$.
    \end{Remark}
    Из~\eqref{eqRealityCondition} и выбора локального параметра следует
    \begin{Statement}
        При четном $\abs{\muld{n}(p)}$ главные коэффициенты $r_j(p)$, $r^+_j(p)$ являются вещественными, а при нечетном~--- чисто мнимыми.
    \end{Statement}

    Нам потребуются некоторые сведения о соотношении знаков главных коэффициентов.
    \begin{Lemma}
        Рассмотрим произвольную вершину $p \in V(\gterm{D})$. Для любого $j = 1, 2, \dots, d$, при котором $\muld{n}_j(p) \ne 0$, выполняется
        \begin{gather}\label{eqPositiveMultOfMainKoef}
            r_j(p) r^+_j(p) > 0
        \end{gather}
    \end{Lemma}
    \begin{Proof}
        Рассмотрим ограничение $h(\lambda)$, $\lambda \in c$, волновой функции $\Psi(p, \lambda)$ на неподвижный овал $c$. По~\eqref{eqRealityCondition} это ограничение является вещественной либо чисто мнимой функцией на окружности, поэтому можно говорить об ее знаке. По замечанию~\ref{remarkLocParamForOtherPoints}, $r^+_j(p)$ является главным коэффициентом разложения $h(\lambda)$ в точке $A^-_j$. Функция $h(\lambda)$ может менять знак только в точках $A^\pm_1, A^\pm_2, \dots, A^\pm_d$, поскольку во всех остальных точках овала $c$ ее значения отличны от нуля и бесконечности. Заметим, что при переходе через точку четного порядка (неважно, полюс это или нуль) знак сохраняется, а при переходе через точку нечетного порядка~--- меняется на противоположный. Рассмотрим два случая.
        \begin{enumerate}
            \item Пусть сначала $\abs{\muld{n}(p)}$ четно, тогда $h(\lambda)$ вещественна. Покажем, что знаки $r_j(p)$ и $r^+_j(p)$ совпадают. Действительно, если $\muld{n}_j(p)$ четно, то сумма порядков во всех остальных выделенных точках также четна, а следовательно, функция $h(\lambda)$ на пути от $A^+_j$ до $A^-_j$ меняет знак четное количество раз. Если же $\muld{n}_j(p)$ нечетно, то и знак $h$ на рассмотренном пути меняется нечетное число раз, что также говорит о совпадении знаков.
            \item Пусть теперь $\abs{\muld{n}(p)}$ нечетно, тогда $h(\lambda)$ чисто мнимая. Аналогичными рассуждениями получаем, что знаки $r_j(p)$ и $r^+_j(p)$ различаются, тогда искомое неравенство будет выполняться засчет их чистой мнимости.
        \end{enumerate}
        Таким образом, в обоих случаях~\eqref{eqPositiveMultOfMainKoef} выполняется.
    \end{Proof}

    Будем говорить, что числа $a \in \mathbb{C}$, $b \in \mathbb{C}$ одного знака и писать $a \sim b$, если $b \ne 0$ их отношение положительно $\frac{a}{b} > 0$.
    \begin{figure}[h]
        \centering
        \includegraphics{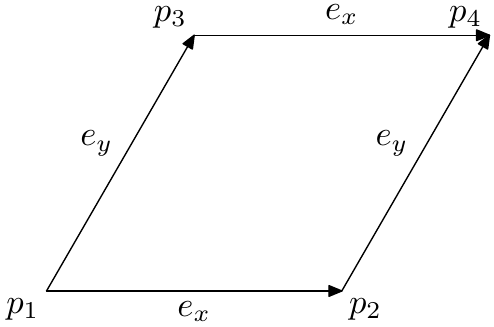}\\
        \caption{Произвольная грань из леммы~\ref{lemmaParallelogramProportion}}
        \label{picGenericFace2}
    \end{figure}
    \begin{Lemma}\label{lemmaParallelogramProportion}
        Рассмотрим произвольную грань $(p_1, p_2, p_4, p_3) \in F(\gterm{D})$. Не умаляя общности будем считать, что ребра этой грани ориентированы в сторону вершин с большими номерами (см. рисунок~\ref{picGenericFace2}). Пусть ребро $(p_1, p_2)$ имеет метку $e_x$, а ребро $(p_1, p_3)$~--- метку $e_y$, $1 \le x, y \le d$, $x \ne y$. Справедливы следующие соотношения
        \begin{gather}\label{eqParallelogramProportion}
            \frac{r_x(p_2)}{r_x(p_4)} \sim \frac{r_x(p_1)}{r_x(p_3)}, \quad \frac{r_y(p_3)}{r_y(p_4)} \sim \frac{r_y(p_1)}{r_y(p_2)}.
        \end{gather}
    \end{Lemma}
    \begin{Proof}
        Мы будем писать $\Psi(p)$ вместо $\Psi(p, \gamma)$ там, где второй параметр понятно какой. Вспомним равенство Коши-Римана
        \begin{gather}\label{eqCRwithMainKoef}
            \Psi(p_4) - \Psi(p_1) = \frac{r_y(p_4)}{r_y(p_3)} \left(\Psi(p_3) - \Psi(p_2)\right) = -\frac{r_x(p_4)}{r_x(p_2)} \left(\Psi(p_3) - \Psi(p_2)\right).
        \end{gather}
        Выражая $\Psi(p_4)$ через остальные слагаемые, вычислим
        $$
            \res{A^+_y} \Psi(p_4) \Psi^+(p_1) \Omega = \frac{r_y(p_4)}{r_y(p_3)} \res{A^+_y} \Psi(p_3) \Psi^+(p_1) \Omega = -\frac{r_y(p_4)}{r_y(p_3)}.
        $$
        Первый переход справедлив, так как только $\Psi(p_3)$ при умножении на $\Psi^+(p_1)$ дает полюс в $A^+_y$. Во втором переходе мы воспользовались тем, что у дифференциала $\Psi(p_3) \Psi^+(p_1) \Omega$ только три полюса: в $R_+$, $R_-$, $A^+_y$, причем вычет в каждом из  $R_+$, $R_-$ равен $\tfrac12$ благодаря соотношению для соседних вершин $\Psi(p_3, R_-) \Psi^+(p_1, R_-) = -1$, откуда вычет в третьем полюсе равен $-1$.

        С другой стороны, у $\Psi(p_4) \Psi^+(p_1) \Omega$ в $A^+_y$ полюс первого порядка, и вычет легко выражается через главные члены разложений в ряды при локальном параметре $z_y$:
        $$
            \res{A^+_y} \Psi(p_4) \Psi^+(p_1) \Omega = r_y(p_4) r^+_y(p_1) \Omega(A^+_y),
        $$
        где под $\Omega(A^+_y)$ имеется в виду значение $f(A^+_y)$ при $\Omega(z_y) = f(z_y)dz_y$ в окрестности $A^+_y$. Из двух этих равенств получаем
        $$
            -\frac{1}{\Omega(A^+_y)} = r_y(p_3)r^+_y(p_1).
        $$

        Аналогичными рассуждениями получаются равенства
        $$
            -\frac{r_x(p_4)}{r_x(p_2)} = \res{A^+_x} \Psi(p_4)\Psi^+(p_1)\Omega = r_x(p_4)r^+_x(p_1)\Omega(A^+_x) \Rightarrow
            -\frac{1}{\Omega(A^+_x)} = r_x(p_2)r^+_x(p_1),
        $$
        $$
            -\frac{r_y(p_3)}{r_y(p_4)} = \res{A^+_y} \Psi(p_3)\Psi^+(p_2)\Omega = r_y(p_3)r^+_y(p_2)\Omega(A^+_y) \Rightarrow
            -\frac{1}{\Omega(A^+_y)} = r_y(p_4)r^+_y(p_2),
        $$
        $$
            -\frac{r_x(p_2)}{r_x(p_4)} = \res{A^+_x} \Psi(p_2)\Psi^+(p_3)\Omega = r_x(p_2)r^+_x(p_3)\Omega(A^+_x) \Rightarrow
            -\frac{1}{\Omega(A^+_x)} = r_x(p_4)r^+_x(p_3).
        $$
        Откуда следуют
        $$
            \frac{r_y(p_3)}{r_y(p_4)} = \frac{r^+_y(p_2)}{r^+_y(p_1)}, \quad \frac{r_x(p_2)}{r_x(p_4)} = \frac{r^+_x(p_3)}{r^+_x(p_1)}.
        $$
        В каждом из них избавимся от $r^+$ с помощью соотношения~\eqref{eqPositiveMultOfMainKoef} и получим искомое~\eqref{eqParallelogramProportion}.
    \end{Proof}

\begin{figure}
      \centering
      \includegraphics[width=300pt]{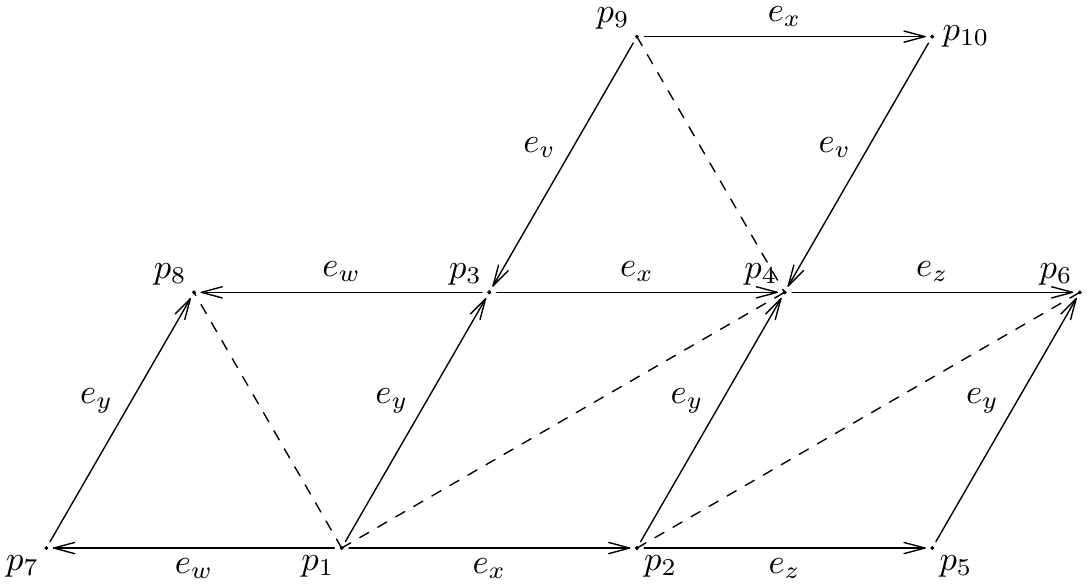}\\
      \caption{Три случая взаимного расположения: $(p_1, p_2, p_5, p_6, p_4, p_3)$, $(p_2, p_1, p_7, p_8, p_3, p_4)$, $(p_1, p_3, p_9, p_{10}, p_4, p_2)$}
      \label{picMutualArrangements}
    \end{figure}

    Итак, благодаря условиям на обобщенные спектральные данные мы получили соотношения, достаточные для выводов о знаках весовой функции. Теперь мы готовы перейти к определению условий на граф $\gterm{D}$, необходимых и достаточных для того, чтобы для любых двух соседних граней графа значения весовой функции на них имеют один знак. Все возможные случаи взаимного расположения с точностью до поворотов и учетом ориентации ребер представлены на рисунке~\ref{picMutualArrangements}.

\begin{figure}[h]
        \centering
        \includegraphics{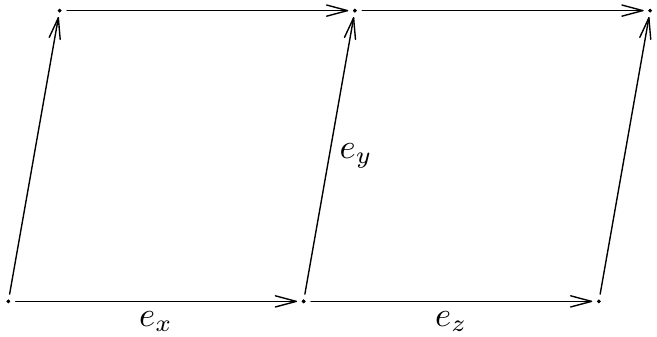}\\
        \caption{$A^+_y$ не должна лежать между $A^+_x$, $A^+_z$}
        \label{picGraphConsistencyRule1}
    \end{figure}

\begin{figure}[h]
        \centering
        \includegraphics{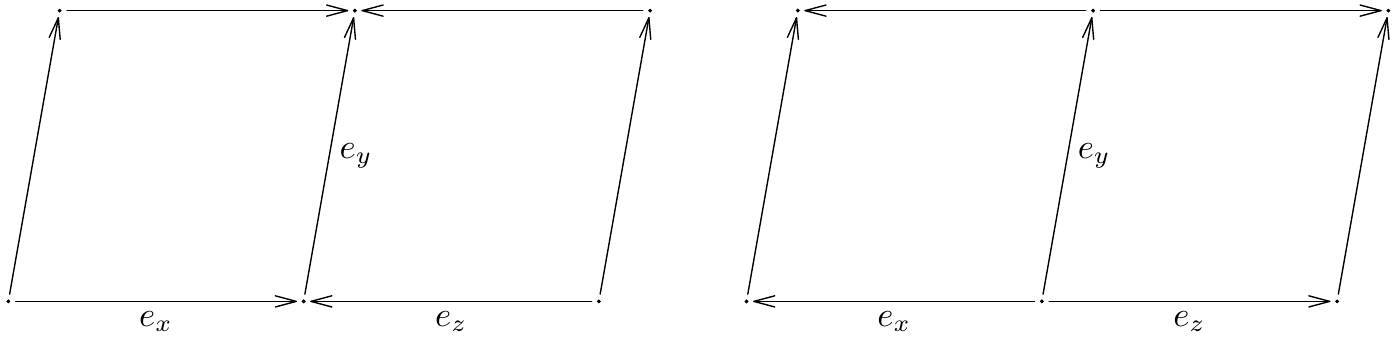}\\
        \caption{$A^+_y$ должна лежать между $A^+_x$, $A^+_z$}
        \label{picGraphConsistencyRule2}
    \end{figure}

    С помощью~\eqref{eqCRwithMainKoef} можно быстро расписать значения весовой функции в терминах главных коэффициентов. При однознаковых преобразованиях мы используем~\eqref{eqParallelogramProportion}.
    $$
        i\nu(p_1, p_4) = \frac{r_y(p_3)}{r_y(p_4)} = -\frac{r_x(p_2)}{r_x(p_4)} \sim -\frac{r_x(p_1)}{r_x(p_3)},
    $$
    $$
        i\nu(p_9, p_4) = -\frac{r_v(p_3)}{r_v(p_4)},
    $$
    $$
        i\nu(p_2, p_6) = -\frac{r_z(p_5)}{r_z(p_6)} \sim -\frac{r_z(p_2)}{r_z(p_4)},
    $$
    $$
        i\nu(p_1, p_8) = \frac{r_w(p_7)}{r_w(p_8)} \sim \frac{r_w(p_1)}{r_w(p_3)}.
    $$
    Разберем требование
    $$
        i\nu(p_1, p_4) \sim i\nu(p_9, p_4) \equivlogic \frac{r_y(p_3)}{r_y(p_4)} \sim -\frac{r_v(p_3)}{r_v(p_4)}.
    $$
    Для этого рассмотрим мероморфную на $\Gamma$ функцию $f(\gamma) = \Psi(p_3, \gamma) / \Psi(p_4, \gamma)$. Ее ограничение на овал $c$ чисто мнимое (вершины $p_3$ и $p_4$~--- соседние), имеет один нуль первого порядка в $A^+_x$ и один полюс первого порядка в $A^-_x$. Нужное требование эквивалентно тому, что значения $f(A^+_y)$ и $f(A^+_v)$ имеют разный знак. Другими словами, точки $A^+_y$ и $A^+_v$ лежат по разные стороны относительно $A^+_x$. Аналогичными рассуждениями с остальными соотношениями приходим к следующим выводам.
    \begin{enumerate}
        \item $i\nu(p_1, p_4) \sim i\nu(p_9, p_4) \equivlogic$ ($A^+_x$ лежит между $A^+_y$ и $A^+_v$).
        \item $i\nu(p_1, p_4) \sim i\nu(p_2, p_6) \equivlogic -\frac{r_x(p_2)}{r_x(p_4)} \sim -\frac{r_z(p_2)}{r_z(p_4)} \equivlogic$ ($A^+_y$ \textit{не} лежит между $A^+_x$, $A^+_z$).
        \item $i\nu(p_1, p_4) \sim i\nu(p_1, p_8) \equivlogic -\frac{r_x(p_1)}{r_x(p_3)} \sim \frac{r_w(p_1)}{r_w(p_3)} \equivlogic$ ($A^+_y$ лежит между $A^+_x$, $A^+_w$).
    \end{enumerate}
    Сформулируем найденное условие.
    \begin{Definition}
        Рассмотрим две произвольные соседние грани квад-графа $(p_1, p_2, p_3, p_4) \in F(\gterm{D})$, $(p_2, p_5, p_6, p_4) \in F(\gterm{D})$. Пусть ребро $(p_2, p_4)$ имеет метку $e_y$, ребро $(p_1, p_2)$~--- метку $e_x$, ребро $(p_2, p_5)$~--- метку $e_z$. Ориентация $(p_2, p_4)$ может быть произвольной. Потребуем, чтобы ребра $(p_1, p_2)$ и $(p_2, p_5)$ были направлены в разные стороны (имели общий конец или общее начало) тогда и только тогда, когда точка $A^+_y$ лежит между $A^+_x$, $A^+_z$ на овале $c$ (см. рисунки~\ref{picGraphConsistencyRule1},~\ref{picGraphConsistencyRule2}). Тогда будем говорить, что разметка ребер квад-графа $\gterm{D}$ положительно согласована с их ориентацией.
    \end{Definition}
    \begin{Example}
        У стандартной квадратной решетки ($d = 2$) координат разметка ребер положительно согласована с их ориентацией по возрастанию координат.
    \end{Example}
    \begin{Example}
        Рассмотрим квадратную решетку, вложенную в $\mathbb{Z}^3$ в виде лесенки: на вертикальных сторонах решетки написано $e_3$, а на горизонтальных чередуются $e_1$ с $e_2$, ориентация ребер такая же. Положительная согласованность также достигается, так как любые две соседних грани отвечают случаю рисунка~\ref{picGraphConsistencyRule1} и порождают условия <<точка $A^+_3$ не лежит между $A^+_1$, $A^+_2$>> и тривиальное <<точки $A^+_1$, $A^+_2$ не лежат между $A^+_3$ и $A^+_3$>>. Первое выполняется по замечанию~\ref{remarkProperPointsEnumeration}.

        Выберем в нашей квадратной решетке вертикальную прямую из ребер и поменяем ориентацию на противоположную у всех горизонтальных ребер слева от этой прямой. В каждой из половинок по отдельности положительная согласованность продолжит выполняться, но на границе появятся противоречия в виде условий <<$A^+_3$ лежит между $A^+_1$, $A^+_2$>>. Положительную согласованность можно вернуть, добавив еще две размерности, $d = 5$, и поменяв пометку каждого горизонтального ребра слева от выбранной прямой: с $e_1$ на $e_4$ и с $e_2$ на $e_5$.
    \end{Example}

    Таким образом, доказана
    \begin{Theorem}[Достаточное условие положительности весовой функции]
        Пусть $\Gamma$ является М-кривой, выделенные точки $A^\pm_j$, $j = 1, 2, \dots, d$, попадают на овал $c$, на остальные овалы попадает по одной точке $\gamma$-дивизора: $\gamma_k \in a_k$, $k = 1, \dots, g$. Все значения весовой функции $\nu$, построенной по этим спектральным данным, имеют один знак тогда и только тогда, когда разметка ребер квад-графа $\gterm{D}$ положительно согласована с их ориентацией.
    \end{Theorem}

\end{document}